# Excluded volume geometry and packing fraction in binary convex hyperparticle mixtures

H.J.H. Brouwers,
*Department of the Built Environment, Eindhoven University of Technology*
*P.O. Box 513, 5600 MB Eindhoven, The Netherlands*
*Email:jos.brouwers@tue.nl*

In this paper the excluded volume of binary similar hyperparticles with small size difference in D-dimensional Euclidean spaces $\mathbb{R}^2$, $\mathbb{R}^3$, and $\mathbb{R}^4$ is studied using two different statistical geometry approaches. These geometric approaches, concerning orientation geometry and integral geometry, yield the excluded volume of particle pairs. The excluded volume of rectangles, based on orientation geometry, in Euclidean space $\mathbb{R}^2$ is used to derive an explicit equation for the bidisperse packing fraction, which is compatible with the expression published previously. Next, the excluded volume of pairs of convex particles in D = 2, 3 and 4, resulting from integral geometry, are presented. These excluded volumes are identical with the specific ones for circles and rectangles (D = 2) and (sphero)cylinders (D = 3), derived by orientation geometry. Furthermore, these orientation geometry-based excluded volumes contain geometric measures: particle volume, surface area, mean curvature and the second quermassintegral. They allow for the derivation of closed-form expressions for the random packing fraction of binary convex similar hyperparticles in Euclidean spaces $\mathbb{R}^2$ , $\mathbb{R}^3$ and $\mathbb{R}^4$.

## 1. INTRODUCTION

The excluded volume between two rigid bodies is the volume inaccessible to the center of one body because of the presence of the other, due to their finite size and impenetrability. This concept of excluded volume plays a central role in the statistical mechanics of fluids, polymers, and other many-particle systems.
The excluded volume concept was introduced by Kuhn in 1934 to study polymeric chains [1]. Onsager demonstrated that a phase transition can be predicted based on two-particle (spherocylinders or cylinders) interactions, represented by the second virial term in an expansion of the free energy of the system [2]. The excluded volume, appearing in the second virial coefficient, is used across physics, chemistry, and materials science as the first quantitative link between microscopic geometry or pair potentials and macroscopic thermodynamic behavior. Besides appearing as van der Waals type corrections to the equation of state, the excluded volume also enters polymer theory (self-avoiding chains) [1], phase transitions [2], colloids chemistry [3], crowding effects in biological systems [4], and the percolation threshold [5].
Excluded volume is also a fundamental tool in analyzing packings, where it connects the packing fraction to the statistically averaged volume a particle excludes to the center of another particle due to the possibility of overlap [6-9]. In [6] and [7] the excluded volume, in combination with the contact number, was used to determine the packing fraction of monosized (sphero)cylinders with large aspect ratio and of hyperspheres, respectively. In [8] the random close packing (RCP) and random loose packing (RLP) fractions of monosized disks and spheres in Euclidean spaces $\mathbb{R}^2$ and $\mathbb{R}^3$, respectively, were derived using statistical excluded volume correlations. In [9] the packing fraction of monosized nonspherical particles was assessed by using excluded volume. Recently the excluded volume was used in the description of the random packing fraction of binary similar hyperspheres with small size ratio [10, 11]. This binary packing can be seen as the most elementary and tractable polydisperse packing.
This problem of particle packing is an ancient puzzle in physics and mathematics that has attracted sustained attention over the last millennia [12, 13]. A packing in D-dimensional Euclidean space $\mathbb{R}^D$ is defined as non-overlapping particles, and the packing faction is the fraction of space $\mathbb{R}^D$ covered by the particles. Particle packing for instance concerns the optimal arrangement of objects in space, and is encountered in mathematics, physics, chemistry, biology engineering and architecture. Such problems form a subject of interest in its own right, providing intriguing challenges, but are also at the heart of many material properties of condensed matter. Researchers have long sought to uncover packing geometries and to use them as a route toward understanding the behavior of liquids and amorphous materials.
Hard particle systems, in particular, provide a powerful model for studying liquid–glass–crystal transitions [12, 13, 14]. Hard particle models are also excellent candidates to model colloidal particles. Furthermore, extending the study to amorphous hyperspheres in higher-dimensional spaces not only deepens our understanding of glass formation in three dimensions but also connects the problem to fields such as signal digitization and error-correcting codes [12, 13]. Yet real-world systems often go beyond the idealization of monosized particles: they may be polydisperse, and packing principles therefore also illuminate the structure and dynamics of colloids, emulsions, biological assemblies, and even protein folding.
In [10], the excluded volume concept of two-particle pairs was combined with the statistically probable combinations of small and large particle pairs, yielding an analytical expression for the random packing fraction of binary similar (sphero)cylinders with small size disparity. This geometric approach of particle packing was successfully validated against a broad collection of computational and experimental data of these packings in three dimensions [9].

The governing analytical expression, that followed from the excluded volume approach, contains a factor (1 – f), where f is the monosized packing fraction of the particle shape concerned. Furthermore, a contraction function v(u) emerged, where u is the size ratio of the two particles in the binary mix. Though the underlying excluded volume expressions of Onsager applied to particle classes (sphero)cylinders only, they were asserted to hold for all particles shapes in three-dimensional Euclidean space (spatial dimension D = 3) [10]. Furthermore, based on analogical reasoning, the approach was extended to the excluded volume in other Euclidean dimensions [10]. In [11] this assertion was confirmed by a successful application of the model by predicting the packing fraction of two-dimensional circle packings (D = 2) and the effect of bidispersity on glass transition of hyperspheres in the infinite spatial dimension limit ($D \rightarrow \infty$).

The use of excluded volume to determine the monosized packing fraction is qualitatively correct only [6, 7, 10]. Notwithstanding this truth for the monodisperse case, the excluded volume approach hence appeared to yield the quantitatively correct expression for the bidisperse case, in particular at the transition between monodisperse and bidisperse packing (size ratio close to unity). This conclusion followed from studying many particle shapes and many spatial dimensions [10, 11]. This *ansatz* is also adopted here.

In this paper we will provide a generalization of the excluded volumes of binary similar convex particles in Euclidean spaces $\mathbb{R}^2$, $\mathbb{R}^3$ and $\mathbb{R}^4$. To this end, the integral geometry approach, introduced by Isihara [15], to the excluded volume of convex particles is employed. Where Onsager determined the excluded volume on the orientationally averaged excluded volume of a particle pair, Isihara [15], Isihara and Hayashida [16, 17], Boublik [18], Torquato and Jiao [19, 20] and Kulossa and Wagner [21], all used Minkowksi [22] integrals. This approach allows for a generalized approach to binary random packings of convex particles in Euclidean spaces $\mathbb{R}^2$, $\mathbb{R}^3$, $\mathbb{R}^4$ and beyond, regardless of the particle shape, and so not limited to (sphero)cylinders in $\mathbb{R}^3$ only.

Here, in Section 2 first the binary packing model based on Onsager's excluded volume is recapitulated. In this section also the random binary packing of rectangles in a plane is modelled. For this the excluded volume (actually, area) as determined by Chatterjee [23] is employed, which is based on an orientation averaged excluded surface area of the two particles. This is an approach similar to Onsager's, which is termed "orientation geometry" here, and it is shown that for this two-dimensional case the previously derived packing fraction expressions for D = 3 [10] are applicable to this two-dimensional particle class indeed.

Next, in Section 3 the excluded volumes in D = 2, 3, and 4 are determined based on integral geometry [15-21]. It follows that the resulting excluded volumes are compatible with the ones provided by Onsager [2], *i.e.* binary (sphero)cylinders in $\mathbb{R}^3$, and Chatterjee [23], *i.e.* binary rectangles in $\mathbb{R}^2$. The resulting expressions for the binary random packing fractions, grounded in this alternative excluded volume theory, also yield the factor (1 – f) and the same contraction functions v(u), derived for D = 3, and postulated for other spatial dimensions, earlier [10]. The conclusions are collected in Section 4.

By bridging insights from orientation and integral geometry, this work contributes to the broader understanding of the excluded volume and binary packing fraction of similar and convex binary hyperparticles with small size difference in disordered systems, in spatial dimensions D = 2, 3 and 4.

## 2. ORIENTATION GEOMETRY

This section recapitulates the orientation geometry approach to excluded volume, as first employed by Onsager [2], and its application to binary particle packing in spatial dimension D = 3. Subsequently, the two-dimensional case of rectangles in a plane is presented. Onsager developed this original geometric model for the isotropic liquid-to-nematic phase transition of hard rodlike (spherocylinders and cylinders) particles, which was published in his seminal paper. Onsager demonstrated that a phase transition can be predicted based on two-particle interactions represented by the second virial term in an expansion of the free energy of the system. Onsager based these expressions on the orientationally averaged excluded volume of two spherocylinders or two cylinders with unequal lengths and diameters.

| **D** | **w(u, D)** | **w(1, D)** |
|---|---|---|
| 2 | 1 | 1 |
| 3 | $u + 1$ | 2 |
| 4 | $u^2 + 10u/7 + 1$ | 24/7 |

Table I Function w(u, D) and its value at u = 1 for various dimensions D [10].

In [10, 11] the resulting equations have been employed to determine the packing fraction of assemblies of binary (discretely sized) similar particles in D-dimensional space, characteristic sizes $d_1$ and $d_2$, their ratio u close to unity, and with a normalized number distribution

$$P(d) = X_1\, \delta(d - d_1) + X_2\, \delta(d - d_2) \quad , \qquad (1)$$

where δ is the Dirac delta function, and $X_1$ and $X_2$ are the number fractions of the two components for which the following identity holds

$$X_1 + X_2 = 1 \quad . \qquad (2)$$

### 2.1 Excluded volume and binary random packing fraction

By employing the excluded volume model of Onsager [2], in [10] the following equation was derived for the random packing fraction of similar binary D-dimensional particles, asserting that mixes and two monodisperse assemblies possess same compaction, and a small size difference:

$$\eta(u, X_1, D) = \frac{f(X_1(u^D - 1) + 1)}{X_1(u^D - 1) + 1 - X_1(1 - X_1)(1 - f)v(u, D)} \quad , \qquad (3)$$

with $\eta(u, X_1, D)$ as the binary random packing fraction, f as the monosized packing fraction, u as size ratio $d_1/d_2$ and $v(u, D) \geq 0$ as the contraction function. The numerator of Eq. (3) denotes the volume of the two particles, and the denominator denotes the volume of the packing, which follows from the excluded volume [10]. Throughout this paper, u is the ratio of some characteristic size of both particles, and hence $u^D$ is the volume ratio (surface area ratio for D = 2).

The last term of the denominator in Eq. (3) is negative, reflecting a packing volume contraction, resulting in a packing increase by bidispersity. This contraction term is governed by the product $X_1(1 - X_1)(1 - f)v(u, D)$, where $X_1(1 - X_1)$ accounts for the composition, (1 – f) for the monosized void fraction (depending on particle shape and densification) and v(u, D) is the contraction function (depending on size ratio and dimension).

The monosized packing fraction depends on particle shape, and on compaction. The compaction determines which configuration is attained between loosest or closest modes of the particle packing. The assembly's density will be situated between RLP and RCP. These random particle packings are prototypical glasses in that they are maximally disordered while simultaneously being mechanically rigid. Moreover, size dispersity frustrates crystallization and is therefore a glass phase enabler. Indeed, the glass transition is related to a specific packing density; in "Table II" [14] packing fraction values for different protocols are listed. In [11], the effect of bidispersity on the packing fraction could be quantitatively related to change in glass transition density in the infinite dimension limit, which was perfectly in agreement with statistical mechanical mean-field theory, based on the replica liquid theory to determine the fluid-glass transition in high-dimensions.

The contraction term also contains the contraction function v(u, D). This contraction function was described as [10]:

$$v(u, D) = w(u, D)\,(u - 1)^2 \quad . \qquad (4)$$

The w(u, D) function is obtained by solving

$$u^D + 1 - 2^{1-D}(u + 1)^D = (1 - 2^{1-D})\, w(u, D)\,(u - 1)^2 \quad . \qquad (5)$$

For spatial dimensions D = 2, 3,… 10, w(u, D) was determined [10], and in Table I they are given for D = 2, 3 and 4. These w(u, D) functions are polynomials in u of order D – 2, with all terms having positive coefficients. The left-hand side of Eq. (5) follows from the scaled contraction term of D-dimensional spheres (hyperspheres) based on excluded volume, and the right-hand side is the asserted (1 – f)v(u, D) function.

The binary excluded volume and contraction function followed from using the Onsager ensemble averaged excluded volume of uneven particle pairs of (sphero)cylinders, and assessing their statistical occurrence in the assembly. These cylinders and spherocylinders are two particle classes, which may each take arbitrary l/d (length/diameter ratios), each l/d constituting a distinct particle shape, having a specific RLP and RCP fraction. But strictly speaking, Eq. (3) was derived and validated (experimentally and computationally), for these two particle classes in $\mathbb{R}^3$ only [10]. Eqs. (4) and (5) are based on the insight in the nature of w(u, D = 3) of these particle classes. Eq. (3) was therefore proposed as a generalized expression, applicable to all particle shapes in arbitrary Euclidean space. In [11], Eq. (3) compared well with simulations concerning RCP of bidisperse circles in dimension D = 2, and with hyperspheres in $D \to \infty$.

Though the Onsager model based contraction function v(u) is accurate near u = 1, for larger size differences the function

$$v(u, D) = \frac{(u^D - 1)^2(1 - D^{-1})}{2(u^D + 1)(1 - 2^{1-D})} \quad , \qquad (6)$$

appeared to be more accurate in the larger size difference range [10, 11]. Eq. (6) is asymptotically identical to Eq. (4) for $u \to 1$, and so towards the limit of a monosized system. But the validity of Eq. (6) stretches to size ratios u deviating more from unity [10, 11].

This convergence for $u \to 1$ can be verified by inserting the asymptotic approximations

$$((1 + \varepsilon)^D - 1)^2 = (\varepsilon D)^2 + O(\varepsilon^3) \quad , \qquad (7)$$

and

$$(1 + \varepsilon)^D + 1 = 2 + \varepsilon D + O(\varepsilon^2) \quad , \qquad (8)$$

into Eq. (6), taking $\varepsilon = u - 1$, so that Eq. (6) is approximated by

$$v(u, D) = \frac{(u - 1)^2 (D^2 - D)}{4(1 - 2^{1-D})} + O((u-1)^3) \quad . \tag{9}$$

For dimensions D = 2, 3 and 4 it can be verified that Eq. (9) corresponds to Eq. (4) invoking the values for w(u = 1, D) of Table I.

Figure 1 shows the two equations, and their convergence for $u^D \downarrow 1$ (so $X_1$ representing the large component number fraction) and D = 2 and 3. The figure displays Onsager based Eq. (4) for D = 2 and 3 (with w(u, D) taken from Table I), and Eq. (6). But for larger size differences (u deviating more from unity), it appeared that Eq. (4) overshoots the effect of bidispersity on the contraction term and, hence, also the binary random packing fraction. As Figure 1 illustrates, for a given volume ratio $u^D$, this overshoot becomes more pronounced with increasing spatial dimension D.

In [10, 11] it was namely seen that for both RCP and RLP, and in $\mathbb{R}^2$ and $\mathbb{R}^3$, the combination of Eq. (3) and (6) is accurate up to $u^D$ of about 9. Eqs. (3) and (6) are also available and accurate up to the infinite dimension limit [11]. Hence, Eq. (6) also has the additional advantage that it provides the contraction function for all spatial dimensions, whereas the Onsager based contraction function was determined for D = 2, 3,…, 10 only [10]. Eqs. (3) and (6) furthermore reveal that the packing increase by binary dispersity is governed by volume ratio $u^D$ (for D = 2 it constitutes the surface area ratio) of the two similar particles.

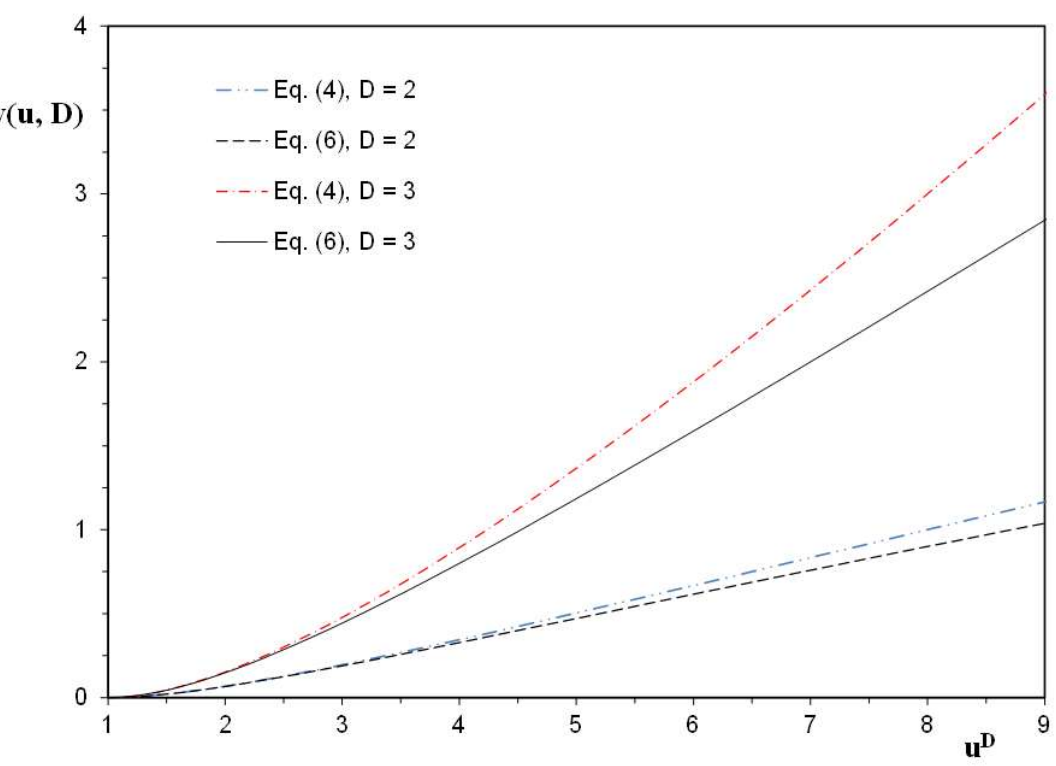

Figure 1 Contraction function v(u, D) for size ratio $u^D$ ranging from unity to 9, and for spatial dimensions D = 2 and 3. Eq. (4), with w(u, D) from Table I, and Eq. (6) are shown.

The binary random packing fraction as governed by Eq. (3), whether Eq. (4) or Eq. (6) is used for v(u, D), is invariant to u smaller or larger than unity. If we use the range u = 1 to ∞, then $X_1$ is the larger one of the two components, if u ranges from 0 to 1, $X_2$ is the larger one. In [10, 11], and here (*e.g.* Figure 1), u ranges from zero to infinity (in [10, 11], $X_1$ was therefore referred to as "$X_L$"). This insight also has as a consequence that the model is accurate for $1/9 \leq u^D \leq 9$.

It is furthermore noteworthy that the proposed model is based solely on geometric considerations (orientation average of the excluded volume of particle pairs, and the statistical occurrence of these pairs), and no adjustable parameters had to be introduced. The governing parameters, *i.e.* the monosized packing fraction f, size ratio u, and concentration $X_1$ are all physically defined.

## 2.2 Rectangles in two dimensions

In this subsection the binary random packing fraction of similar rectangles in two dimensions (*i.e.* the Euclidean plane) is determined. This packing fraction will again be based on their excluded volume and again using orientation geometry.

Packing problems in two dimensions are central to a wide range of disciplines, including statistical physics, materials science, and computational geometry. In physics the packing of two-dimensional objects on a plane is studied as an introduction to the three-dimensional (D = 3) problem. Furthermore, the packing in a plane has been used to model the structure of monolayer films, the adsorption on substrates, percolation, and the organization of cells [5, 12, 18, 23, 24].

While the dense packing of symmetric shapes such as monodisperse [24] and binary disks [11] has been extensively studied, the packing of anisotropic particles, such as rectangles, introduces additional complexity due to their direction-dependent interactions. These systems exhibit rich structural and dynamical behaviors and are relevant to the modeling of composite materials, liquid crystals, and wireless networks.

The orientationally averaged excluded volume of binary rectangles was derived by Balberg *et al.* [5], Li and Östling [25] and Chatterjee [23] for arbitrary l/d (length/width ratio), whereby the expression provided by [23] is correct [26, 27]. To derive the binary random packing fraction of this particle class, a similar procedure is followed as for determining the binary random packing fraction of (sphero)cylinders [10].

We study an assembly of two rectangles with the same shape, governed by l/d, where d and l are the width and length of the rectangles. As said, the rectangles may take any l/d, each l/d being a distinct particle shape, having a specific monosized packing fraction. And, obviously, it is sufficient to consider either l/d ≥ 1 or l/d ≤ 1 to cover all conceivable rectangle shapes.

Regardless of the rectangle shape, in view of similarity, all dimensions have an identical size ratio u, so

$$\frac{d_1}{d_2} = \frac{l_1}{l_2} = u \quad . \tag{10}$$

The surface areas of both particles (i = 1 or 2) and their ratio read:

$$V_{pi} = d_i l_i ; \; V_{p1} = u^2 V_{p2} \quad . \tag{11}$$

The convex addition of the rectangles' surface area of two rectangles reads

$$2V_p = 2X_1V_{p1} + 2X_2V_{p2} = 2V_{p2}(1+(u^2-1)X_1) \quad , \qquad (12)$$

whereby Eqs. (2) and (11) have been inserted.
The ensemble average of all excluded areas of two binary rectangles reads [23]:

$$V_e^{i,j} = (l_i d_i + l_j d_j) + \frac{2(l_i l_j + d_i d_j + l_i d_j + l_j d_i)}{\pi} \quad . \qquad (13)$$

As already introduced in [10] for binary cylinders and spherocylinders, the mean excluded area of randomly mixing the small and large rectangles follows from the statistically probable combinations of these rectangles:

$$V_e = X_1^2 V_e^{1,1} + X_2^2 V_e^{2,2} + X_1 X_2 (V_e^{1,2} + V_e^{2,1}) \quad . \qquad (14)$$

which is

$$V_e = X_1 V_e^{1,1} + X_2 V_e^{2,2} - X_1 X_2 (V_e^{1,1} + V_e^{2,2} - 2V_e^{1,2}) , \qquad (15)$$

in view of

$$V_e^{2,1} = V_e^{1,2} \quad , \qquad (16)$$

and

$$X_i^2 = X_i - (1 - X_i)X_i \quad . \qquad (17)$$

Furthermore, following Eqs. (10), (11) and (13)

$$V_e^{2,2} = 2V_{p2} + \frac{2(l_2^2 + d_2^2 + 2V_{p2})}{\pi} \quad , \qquad (18)$$

$$V_e^{1,1} = u^2 V_e^{2,2} \quad , \qquad (19)$$

$$V_e^{1,2} = V_{p2}(u^2 + 1) + \frac{2u(l_2^2 + d_2^2 + 2V_{p2})}{\pi} \quad , \qquad (20)$$

so that, with Eqs. (2) and (12), Eq. (15) becomes

$$V_e = 2V_e^{2,2}(1+(u^2-1)X_1) - 2X_1(1-X_1)(V_e^{2,2} - 2V_{p2})(u-1)^2 . \qquad (21)$$

For $X_i = 1$ ($i = 1$ or 2), the monosized cases, yield

$$f = \frac{2V_{pi}}{V_e^{i,i}} = \frac{\pi l_i d_i}{l_i^2 + d_i^2 + (2+\pi) l_i d_i} \quad , \qquad (22)$$

see Eqs. (11) and (13). For the binary case holds:

$$\eta = \frac{2V_p}{V_e} \quad . \qquad (23)$$

Substituting Eqs. (12) and (21) yields

$$\eta(u, X_1, 2) = \qquad (24)$$

$$\frac{2V_{p2}(1+(u^2-1)X_1)}{V_e^{2,2}(1+(u^2-1)X_1) - X_1(1-X_1)(V_e^{2,2} - 2V_{p2})(u-1)^2} \quad .$$

Dividing the numerator and dominator by $V_e^{2,2}$ and substituting Eq. (22) results in Eqs. (3) and (4), with w(u, 2) from Table I. Hence, this equation is the binary random packing fraction of rectangles, with arbitrary l/d, where each l/d represents a distinct particle shape.
This result supports the postulation formulated in [10] that Eq. (3) is applicable for binary particles in $\mathbb{R}^2$, which was originally based on the binary (sphero)cylinders packing fraction expression in $\mathbb{R}^3$. This postulation led to the factor $(1 - f)(u - 1)^2$ [10], this was validated for circles [11], and now also appears to capture the contraction of the considered binary rectangles packings.

### 3. INTEGRAL GEOMETRY

A powerful geometric framework to describe shape, curvature, surface area, and volume of particles is provided by integral geometry, in particular Minkowski functionals (also called intrinsic volumes, quermassintegrals, *etc.*) [15-22]. Minkowski functionals play a crucial role in integral geometry, a mathematical discipline aiming for a geometric description of objects using integral quantities instead of differential expressions. These give functionals of a particle such as its volume, surface area, mean (integrated) curvature, and Euler characteristic. The excluded volume between two particles follows from their relative positions and orientations. Using integral geometry, the average excluded volume (for random orientations) for convex particles can be expressed in terms of their intrinsic volume, surface area, and mean curvature. In other words, Minkowski integrals provide a way to express geometric invariants relevant to excluded volume formulas [15-22]. For convex hard particles, the excluded volume can be expressed exactly in terms of certain Minkowski functionals (volume, surface area, and mean radius of curvature). Convex shape means that for any two points inside or on the surface of the particle, the straight line connecting those two points lies entirely within the particle. The pioneering work of Isihara [15] and Isihara and Hayashida [16, 17] provided such formulas for rigid convex molecules ("ovaloids"), giving analytic expressions for excluded volume for particles of given shape in terms of volume, surface, and mean curvature (or some averaged radius). Kihara [28] further analyzed Isihara-Hayashida's theory to refine or check approximate formulas.
Later, in two dimensions, Boublík [18] derived the excluded surface area in terms of geometric measures of two particle pairs in a plane. In particular, the excluded area of two convex particles in random orientation was expressed via simple shape invariants: area and perimeter. Also their average excluded area is determined entirely by Minkowski functionals. In four dimensions and beyond, Torquato and Jiao [19, 20] and Kulossa and Wagner [21] provided expressions for the excluded volumes based on integral geometry.
In this section, this integral approach is applied to determine the excluded volumes and the random packing fraction of binary packings in Euclidean spaces $\mathbb{R}^2$, $\mathbb{R}^3$ and $\mathbb{R}^4$.

### 3.1 Two dimensions

The excluded area for convex particles in a plane, using Minkowski integrals, was determined by Boublik [18]. In two dimensions, it appeared that two geometric measures suffice to calculate the excluded area $V_e^{i,j}$ of a pair of particles with surface areas $V_{pi}$ and $V_{pj}$

$$V_e^{i,j} = V_{pi} + V_{pj} + \frac{S_{pi}S_{pj}}{2\pi} \quad , \tag{25}$$

where $S_{pi}$ and $S_{pj}$ are the perimeter of particle i and j, respectively. The excluded volume of two circles with $d_1$ and $d_2$, *viz.* $V_e^{1,2} = \pi(d_1 + d_2)^2/4$, which is the most simple case, readily follows from Eq. (25) and the areas $(\pi d_i^2/4)$ and perimeters $(\pi d_i)$ of both circles (i = 1 and 2).
Also, for rectangles, substituting $V_{pi} = l_i d_i$ and $S_{pi} = 2(l_i + d_i)$, yields Eq. (13). So not surprisingly, orientation and integral geometric approaches lead to the same excluded area for binary circles and rectangles. But the geometric integral approach allows for a derivation of Eqs. (3) and (4) for all convex particles in Euclidean space $\mathbb{R}^2$.
Using Eqs. (2), (11), (15), and (25) yields

$$V_e = 2V_e^{2,2}(1 + (u^2 - 1)X_1) - \frac{X_1(1 - X_1)(S_{p1}^2 + S_{p2}^2 - 2S_{p1}S_{p2})}{2\pi} \quad . \tag{26}$$

For similar particles in two dimensions the following holds:

$$S_{p1} = uS_{p2} \quad , \tag{27}$$

so that

$$S_{p1}^2 + S_{p2}^2 - 2S_{p1}S_{p2} = S_{p2}^2(u-1)^2 \quad . \tag{28}$$

Substituting Eqs. (12), (26), and (28) in Eq. (23) yields

$$\eta(u, X_1, 2) = \frac{2V_{p2}(1 + (u^2 - 1)X_1)}{V_e^{2,2}(1 + (u^2 - 1)X_1) - X_1(1 - X_1)S_{p2}^2(u - 1)^2} \quad . \tag{29}$$

The monosized packing is found for $X_i = 1$ (i = 1 or 2):

$$f = \frac{2V_{pi}}{V_e^{i,i}} = \frac{2V_{pi}}{\left(2V_{pi} + \frac{S_{pi}^2}{2\pi}\right)} \quad . \tag{30}$$

Dividing the numerator and dominator of Eq. (29) by $V_e^{2,2}$ and substituting Eq. (30) results in Eqs. (3) and (4), with w(u, 2) from Table I.
This is the expression for packing of binary particles in a plane that was proposed in [10], and applied to binary circles [11] and rectangles (previous subsection). The derivation in this subsection confirms that the excluded volumes that resulted in Eqs. (3) and (4) are identical for all convex and similar binary particles with small size difference in two dimensions. In other words, they are not restricted to circles and rectangles only.

### 3.2 Three dimensions

The convex addition of the three-dimensional and similar particles' volume reads

$$2V_p = 2X_1V_{p1} + 2X_2V_{p2} = 2V_{p2}(1 + (u^3 - 1)X_1) \quad , \tag{31}$$

whereby Eq. (2) and

$$V_{p1} = u^3V_{p2} \quad , \tag{32}$$

are inserted.

| **Shape** | $V_p$ | $S_p$ | $M_p$ |
|---|---|---|---|
| Sphere<br>diameter d | $\frac{\pi d^3}{6}$ | $\pi d^2$ | $2\pi d$ |
| Cylinder<br>length l, diameter d | $\frac{\pi d^2 l}{4}$ | $\frac{\pi d(d+2l)}{2}$ | $\frac{\pi(\pi d+2l)}{2}$ |
| Spherocylinder<br>length l, diameter d | $\frac{\pi d^2(2d+3l)}{12}$ | $\pi d(d+l)$ | $\pi(2d+l)$ |

Table II Volumes, surface areas, and radii of mean curvature for some randomly oriented convex particles in $\mathbb{R}^3$ [19, 28].

Isihara [15] and Isihara and Hayashida [16, 17] provided excluded volume formulas for rigid convex particles in three dimensions in terms of volume, surface, and mean curvature (some averaged radius):

$$V_e^{i,j} = V_{pi} + V_{pj} + \frac{S_{pi}M_{pj} + S_{pj}M_{pi}}{4\pi} \quad , \tag{33}$$

where $V_{pi}$ and $V_{pj}$ are the volume of particle i and j, respectively; $S_{pi}$ and $S_{pj}$ are their surface areas; and $M_{pi}$ and $M_{pj}$ are their mean curvatures.

For a number of particle shapes the values of $V_p$, $S_p$ and $M_p$ were computed [15-17, 19, 28, 29], and a selection is given in Table II. Note that in [19-21] a slightly different definition of the curvature was introduced ("$\overline{R}$" and "$\tilde{R}_P$", corresponding to $M_p/4\pi$).

Not surprisingly, invoking the $V_p$, $S_p$ and $M_p$ from Table II pertaining to cylinders and spherocylinders in Eq. (33) yields the same excluded volume as obtained by Onsager, which is based on orientation geometry [2, 10].

For similar particles in D = 3 the particles' surfaces and curvatures scale as

$$S_{p1} = u^2 S_{p2} \quad ; M_{p1} = u M_{p2} \quad , \tag{34}$$

so that, with Eq. (32), Eq. (33) yield

$$V_e^{1,1} = u^3 V_e^{2,2} \quad . \tag{35}$$

Eqs. (2), (14), (17), (33) and (35) yields

$$V_e = 2V_e^{2,2}(1+(u^3-1)X_1) - \frac{X_1 X_2 (M_{p1}S_{p1} + M_{p2}S_{p2} - M_{p1}S_{p2} - M_{p2}S_{p1})}{2\pi} \quad . \tag{36}$$

Using Eq. (34) it follows that

$$M_{p1}S_{p1} + M_{p2}S_{p2} - M_{p1}S_{p2} - M_{p2}S_{p1} = M_{p2}S_{p2}(u+1)(u-1)^2 \quad . \tag{37}$$

For $X_i = 1$ (i = 1 or 2), the monosized packing reads

$$f = \frac{2V_{pi}}{V_e^{i,i}} = \frac{2V_{pi}}{\left(2V_{pi} + \frac{M_{pi}S_{pi}}{2\pi}\right)} \quad . \tag{38}$$

Substituting Eqs (31), (36), and (37) in Eq. (23) yields

$$\eta(u, X_1, 3) = \frac{4\pi V_{p2}(1+(u^3-1)X_1)}{2\pi V_e^{2,2}(1+(u^3-1)X_1) - X_1(1-X_1)M_{p2}S_{p2}(u+1)(u-1)^2} . \tag{39}$$

Dividing the numerator and denominator by $2\pi V_e^{2,2}$ and substituting Eqs. (38) results in Eqs. (3) and (4), with w(u, 3) from Table I. This is the expression for random packing of binary particles in three dimensions that was derived in [10], based on the excluded volumes to (sphero)cylinders provided by Onsager using orientation geometry [2].

As (sphero)cylinders are convex particles, it is therefore very probable that Eqs. (3) and (4) are valid for binary mixes of all similar convex particles with small size difference in three dimensions.

### 3.3 Four dimensions

The convex addition of the similar particles' volume in four dimensions reads

$$2V_p = 2X_1 V_{p1} + 2X_2 V_{p2} = 2V_{p2}(1+(u^4-1)X_1) \, , \tag{40}$$

whereby Eq. (2) and

$$V_{p1} = u^4 V_{p2} \quad , \tag{41}$$

are inserted.

For hyperparticles, quermassintegrals enter the expressions for the excluded volume [19-21]. In four dimensions, the excluded volume of a pair of particles is expressed in particles' volume, surface, mean curvature and the second quermassintegral [19-21, 30]:

$$V_e^{i,j} = V_{pi} + V_{pj} + \frac{S_{pi}M_{pj} + S_{pj}M_{pi}}{4\pi} + \frac{12 W_{2i} W_{2j}}{\pi^2} \quad , \tag{42}$$

where $V_{pi}$ and $V_{pj}$ are the volume of particle i and j, respectively; $S_{pi}$ and $S_{pj}$ are their surface areas; $M_{pi}$ and $M_{pj}$ are their mean curvatures; and $W_{2i}$ and $W_{2j}$ are their second quermassintegrals.

| **Shape** | $V_p$ | $S_p$ | $M_p$ | $W_2$ |
|---|---|---|---|---|
| Hypersphere diameter d | $\frac{\pi^2 d^4}{32}$ | $\frac{\pi^2 d^3}{4}$ | $2\pi d$ | $\frac{\pi^2 d^2}{8}$ |
| Hypercylinder length l diameter d | $\frac{\pi d^3 l}{6}$ | $\frac{\pi d^2(d+3l)}{3}$ | $\frac{8(2d+l)}{3}$ | $\frac{\pi d(\pi d+4l)}{12}$ |
| Hyperspherocylinder length l diameter d | $\frac{\pi d^3(3\pi d+16l)}{96}$ | $\frac{\pi d^2(\pi d+4l)}{4}$ | $\frac{2(3\pi d+4l)}{3}$ | $\frac{\pi d(3\pi d+8l)}{24}$ |

Table III Volumes, surface areas, radii of mean curvature, and second quermassintegrals for some randomly oriented convex particles in Euclidean space $\mathbb{R}^4$ [21].

For a number of convex particle shapes the values of $V_p$, $S_p$, $M_p$ and $W_2$ were computed [21], and a selection is given in Table III. Note that in [21] the hyperspherocylinders were characterized by the ratio of the total length (including the two caps) to the diameter, referred to as "γ", which is $l/d+1$.

For $X_i = 1$ (i = 1 or 2), the monosized packing fraction reads

$$f = \frac{2V_{pi}}{V_e^{i,i}} = \frac{2V_{pi}}{\left(2V_{pi} + \dfrac{S_{pi}M_{pi}}{2\pi} + \dfrac{12W_{2i}^2}{\pi^2}\right)} \quad . \tag{43}$$

For similar particles the following holds:

$$S_{p1} = u^3 S_{p2} \ ; \ W_{21} = u^2 W_{22} \ ; \ M_{p1} = u M_{p2} \quad , \tag{44}$$

so that, with Eq. (42),

$$V_e^{1,1} = u^4 V_e^{2,2} \quad . \tag{45}$$

Now, Eqs. (2), (15), (42), and (45) yield

$$\begin{aligned} V_e = {} & 2V_e^{2,2}\left(1+(u^4-1)X_1\right) - \\ & X_1X_2\left(\frac{M_{p1}S_{p1}+M_{p2}S_{p2}}{2\pi} + \frac{12(W_{21}^2+W_{22}^2)}{\pi^2}\right) + \\ & X_1X_2\left(\frac{M_{p1}S_{p2}+M_{p2}S_{p1}}{2\pi} + \frac{24W_{21}W_{22}}{\pi^2}\right) \quad . \end{aligned} \tag{46}$$

Using Eq. (44), the last two terms yield

$$\begin{aligned} & \frac{M_{p1}S_{p1}+M_{p2}S_{p2}}{2\pi} + \frac{12(W_{21}^2+W_{22}^2)}{\pi^2} \\ & -\frac{M_{p1}S_{p2}+M_{p2}S_{p1}}{2\pi} - \frac{24W_{21}W_{22}}{\pi^2} = \\ & \left(\frac{M_{p2}S_{p2}(u^2+u+1)}{2\pi} + \frac{12W_{22}^2(u+1)^2}{\pi^2}\right)(u-1)^2 \quad . \end{aligned} \tag{47}$$

This result reveals that for hyperparticles in general, the contraction term cannot be written in the form of Eq. (4), with the exception of the special case of hyperspheres. The reason is the combination of particle surface, mean curvature, and second quermassintegral in the excluded volume, which is not the case in $\mathbb{R}^2$ and $\mathbb{R}^3$.

As said, for hyperspheres, on the other hand, Eq. (4) applies. Substituting the $S_{p2}$, $M_{p2}$, and $W_{22}$ pertaining to hyperspheres from Table III produces, namely,

$$\begin{aligned} & \left(\frac{M_{p2}S_{p2}(u^2+u+1)}{2\pi} + \frac{12W_{22}^2(u+1)^2}{\pi^2}\right)(u-1)^2 = \\ & \qquad \frac{7\pi^2 d_2^4 (u^3+10u/7+1)(u-1)^2}{16} \quad . \end{aligned} \tag{48}$$

On the right-hand side, one can recognize Eq. (4), with the w(u, 4) function given in Table I. Furthermore, $7\pi^2 d_2^4/16$ can be replaced by $V_e^{2,2} - 2V_{p2}$, since $V_e^{2,2} = \pi^2 d_2^4/2$ and $V_{p2} = \pi^2 d_2^4/32$ (Table III). $V_e^{2,2}$ follows from Table III and Eq. (42) by inserting $d = d_2$, but can also be computed directly by taking $V_p$ from Table III and substituting $d = 2d_2$. Substituting Eqs. (40), (46), (47) and (48) in Eq. (23) yields as binary random packing of hyperspheres in four dimensions

$$\eta(u, X_1, 4) = \frac{2V_{p2}\left(1+(u^4-1)X_1\right)}{V_e^{2,2}\left(1+(u^4-1)X_1\right) - X_1(1-X_1)\left(V_e^{2,2}-2V_{p2}\right)v(u,4)} \ . \tag{49}$$

Dividing the numerator and denominator by $V_e^{2,2}$ and substituting Eq. (43) results in Eqs. (3) and (4). This is the expression for the random packing of binary particles in hyperdimensions that was proposed in [10], and validated for hyperspheres in the infinite dimension limit [11].

## 4. CONCLUSIONS

This paper addresses the excluded volume of similar binary hyperparticles with small size difference, which is instrumental in assessing their packing fraction [10, 11]. This excluded volume is obtained using statistical geometry, that is ensemble averaging of geometric properties. As with atoms, it is difficult or even impossible to track every particle in a real packing. Instead, here we describe it statistically.

Onsager [2] introduced excluded volume as a statistical-mechanical concept to describe orientation entropy of (sphero)cylinders, here termed "orientation geometry", while Isihara [15] developed integral geometry to derive and generalize the excluded volume rigorously from convex-body geometry.

Onsager [2] determined the excluded volumes of three-dimensional (sphero)cylinders by considering all orientations and computing their average. In [10] a general equation for the binary random packing fraction of these three-dimensional particles was derived, Eqs. (3) and (4), which were experimentally and computationally validated. They are based on the statistically averaged excluded volumes of binary (sphero)cylinders with arbitrary length-diameter ratio, combined with the statistically probable combinations of small and large (sphero)cylinders in the assembly. In [11] it was shown that these $\mathbb{R}^3$ expressions [10] are applicable to binary disks in a plane (D = 2) and hyperspheres in the infinite large dimension limit.

Here it is shown that Eqs. (3) and (4) are also compatible with the excluded volume, and hence the binary random packing fraction, of rectangles in a plane with arbitrary length-width ratios. This packing expression is based on the excluded area of rectangle pairs, determined by orientation geometry as well [23]. The orientation geometry approach thus confirms the validity of Eqs. (3) and (4) for several distinct particle classes, *viz.* rectangles and (sphero)cylinders, in Euclidean spaces $\mathbb{R}^2$ and $\mathbb{R}^3$, respectively.

A more general method for validating Eqs. (3) and (4) is offered by assessing the excluded volume by aforementioned integral geometry, based on Minkowski integrals. For spatial dimension D = 2 [18], 3 [15-17, 28, 29] and 4 [19-21], expressions for the excluded area/volume of convex particles of arbitrary shape are provided. These expressions contain the geometric measures: particle volume, surface area, mean curvature, and the second quermassintegral. For assemblies of two similar particles, their ratios are known and the excluded volume can be assessed without the need to determine the geometric measures of each particle shape. Using integral geometry, in both spatial dimensions D = 2 and 3, here it is mathematically proven that the excluded volume that underlies Eqs. (3) and (4) holds for all binary similar convex particles, and so not for circles and rectangles (D = 2) and (sphero)cylinders (D = 3) only. For hyperparticles in spatial dimension D = 4, Eqs. (3) and (4) are valid for the hypersphere only. Probably this conclusion is also valid for higher dimensions. It was namely seen before that for hyperspheres in the infinite dimension limit, Eqs. (3) and (6) appeared to be applicable too [11].

Eq. (6) was proposed as an alternative for Eq. (4) [10]. For size ratio $u \rightarrow 1$, the monosized particle limits, Eqs. (4) and (6) converge (for example see Figure 1 for two and three dimensions). But the alternative contraction function Eq. (6) is accurate for larger size disparities, and available for all spatial dimensions, so from D = 2 to the aforementioned $\infty$ [11].

Finally, it is noteworthy that the description of the binary packing fraction using excluded volume is based solely on physical principles, and does not involve empiricism or calibration factors to obtain the closed-form expressions. The governing parameters, *i.e.* the monosized packing fraction f, size ratio u, number fraction $X_1$, and space dimension D, are all physically defined.

The considered binary packing can be seen as the most elementary and tractable polydisperse packing. The underlying excluded volume approach offers the possibility to statistically model the effect of more complex polydispersity on packing fraction, percolation threshold, glass transition, *etc*.

**ACKNOWLEDGEMENTS**

Dr. A.P. Chatterjee, SUNY College of Environmental Science and Forestry (USA), Dr. I. Balberg, The Hebrew University (Israel), and Dr. A. Wagner of Universität Rostock (Germany) are acknowledged for the discussions.

**DATA AVAILABILITY**

The original contributions presented in this study are included in the article. Any data that are not publicly available are available from the author upon reasonable request.

---

[1] W. Kuhn, Über die Gestalt fadenförmiger Moleküle in Lösungen, Kolloid-Z. 68, 2 (1934) (in German).

[2] L. Onsager, The effects of shape on the interaction of colloidal particles, Ann. N.Y. Acad. Sci. 51, 627 (1949).

[3] P.G. Bolhuis, A.A. Louis J.P. Hansen, Influence of polymer-excluded volume on the phase-behavior of colloid-polymer mixtures, Phys. Rev. Lett. 89, 128302 (2002).

[4] R.J. Ellis, Macromolecular crowding: obvious but underappreciated, Trends in Biochemical Sciences 26, 597 (2001).

[5] I. Balberg, C.H. Anderson, S. Alexander and N. Wagner. Excluded volume and its relation to the onset of percolation. Phys. Rev. B 30, 3933 (1984).

[6] A.P. Philipse, The random contact equation and its implications for (colloidal) rods in packings, suspensions, and anisotropic powders, Langmuir 12, 1127, *ibid.* 5971 (1996).

[7] A.P. Philipse, Caging effects in amorphous hard-sphere solids, Colloids Surf. A 213, 167 (2003).

[8] A. Zaccone Explicit analytical solution for random close packing in d = 2 and d = 3, Phys. Rev. Lett. 128, 028002, *ibid.* 129, 039901(E) (2022). See also discussion arXiv:2201.07629, arXiv:2201.10951, arXiv:2201. 10550, arXiv:2201.12784, arXiv:2204.13901, arXiv:2205.00720, arXiv:2201.06129, arXiv:2202. 05078.

[9] W.X. Xu, Z.W. Ma, J.L. Fu and Y. Jiao, Percolation-based mean field theory for disordered particle packings, Powder Technology 460, 121088 (2025).

[10] H.J.H. Brouwers, Physics – Uspekhi, Random packing fraction of binary similar particles: Onsager's excluded volume model revisited, 67, 510 (2023).

[11] H.J.H. Brouwers, Random packing fraction of binary hyperspheres with small or large size difference: A geometric approach, Phys. Rev. E 112, 025411 (2025).

[12] S. Torquato and F.H. Stillinger, Jammed hard-sphere packings: from Kepler to Bernal and beyond, Rev. Mod. Phys. 82, 2633 (2010).

[13] P.K. Morse and P. Charbonneau, Amorphous Packings of Spheres, in *Packing Problems in Soft Matter Phyics*, pp. 111-126 (eds. H.K. Chan, S. Hutzler, A. Mughal, C.S. O'Hern, Y.J. Wang and D. Weaire), Royal Soc. Chem., Cambridge, UK (2025).

[14] G. Parisi and F. Zamponi, Mean-field theory of hard sphere glasses and jamming, Rev. Mod. Phys. 82, 789 (2010).

[15] A. Isihara, Determination of molecular shape by osmotic measurement, J. Chem. Phys. 18, 1446 (1950).

[16] A. Isihara and T. Hayashida, Theory of high polymer solutions. I. Second virial coefficient for rigid ovaloids model., J. Phys. Soc. Japan 6, 40 (1951).

[17] A. Isihara and T. Hayashida, Theory of high polymer solutions. II. Special forms of second osmotic coefficient, J. Phys. Soc. Japan 6, 46 (1951).

[18] T. Boublik, Two-dimensional convex particle liquid, Mol. Phys. 29, 421 (1975).

[19] S. Torquato and Y. Jiao, Effect of dimensionality on the percolation threshold of overlapping nonspherical hyperparticles, Phys. Rev. E. 87, 022111 (2013).

[20] S. Torquato and Y. Jiao, Exclusion volumes of convex bodies in high space dimensions: applications to virial coefficients and continuum percolation, J. Stat. Phys., 093404 (2022).

[21] M. Kulossa and J. Wagner, Geometric measures of uniaxial solids of revolution in higher-dimensional Euclidean spaces and their relation to the second virial coefficient, Phys. Rev. E 111, 024111 (2025).

[22] H. Minkowski, Volumen und Oberfläche, Math. Ann. 57, 447 (1903) (in German).

[23] A.P. Chatterjee, Percolation thresholds and excluded area for penetrable rectangles in two dimensions, J. Stat. Phys. 158, 248 (2015).

[24] H.J.H. Brouwers, A geometric probabilistic approach to random packing of hard disks in a plane, Soft Matter 19, 8465 (2023).

[25] J. Li and M. Östling. Percolation thresholds of two-dimensional continuum systems of rectangles. Phys. Rev. E 88, 012101 (2013).

[26] I. Balberg, Private communications (2025).

[27] A.P. Chatterjee, Private communications (2025).

[28] T. Kihara, Virial coefficients and models of molecules of gases, Rev. Mod. Phys. 25, 831 (1953).

[29] T. Kihara, On Isihara-Hayashida's theory of the second virial coefficient for rigid convex molecules, J. Phys. Soc. Japan 8, 686 (1953).

[30] J. Wagner, Private communications (2025).